\documentstyle[aps,epsf,multicol,amssymb]{revtex}

\def\breakon{\end{multicols}\widetext\vspace{-.2cm}
\noindent\rule{.48\linewidth}{.3mm}\rule{.3mm}{.3cm}\vspace{.0cm}}

\def\breakoff{\vspace{-.2cm}
\noindent
\rule{.52\linewidth}{.0mm}\rule[-.27cm]{.3mm}{.3cm}\rule{.48\linewidth}{.3mm}
\vspace{-.3cm}
\begin{multicols}{2}
\narrowtext}

\def\bj{{\bf j}}
\def\br{{\bf r}}

\def\bA{{\bf A}}

\def\bH{{\bf H}}

\def\cH{{\cal H}}

\def\cP{{\cal P}}
\def\cT{{\cal T}}

\def\cX{{\cal X}}
\def\cZ{{\cal Z}}

\def\be{\begin{eqnarray}}
\def\ee{\end{eqnarray}}
\def\eq{&=&}
\def\nn{\nonumber}

\def\sbf{\sigma^{\sc bf}}
\def\sph{\sigma^{\sc ph}}

\def\scc{\sigma^{\sc cc}}
\def\Tr{{\,\rm Str\,}}

\begin{document}
\draft
\title{Phase Coherence Phenomena in Superconducting Films}
\author{Julia S. Meyer$^{1,2}$ and B. D. Simons$^1$}
\address{$^1$ Cavendish Laboratory, Madingley Road, Cambridge CB3 0HE,
  UK\\ $^2$ Institut f\"ur Theoretische Physik, Universit\"at zu
  K\"oln, 50937 K\"oln, Germany}
\date{\today}
\maketitle
\begin{abstract}
Superconducting films subject to an in-plane magnetic field exhibit a 
gapless superconducting phase. We explore the quasi-particle spectral properties of the gapless phase and comment on the transport properties. Of particular interest is the 
sensitivity of the quantum interference phenomena in this phase to the nature 
of the impurity scattering. We find that films subject to columnar defects exhibit
a `Berry-Robnik' symmetry which changes the fundamental properties of the 
system. Furthermore, we explore the integrity of the gapped phase. As in 
the magnetic impurity system, we show that optimal fluctuations of the 
random impurity potential conspire with the in-plane magnetic field to 
induce a band of localized sub-gap states. Finally, we investigate the interplay of the proximity effect and gapless superconductivity in thin normal metal-superconductor bi-layers.
\end{abstract}

\begin{multicols}{2}
\narrowtext

\section{Introduction}\label{sec-intro}

It is well established both theoretically and experimentally that bulk 
s-wave superconductivity is robust with respect to the addition of 
non-magnetic impurities (Anderson theorem~\cite{anderson}). By contrast,
magnetic impurities affect a mechanism of pair-breaking on the system 
which suppresses and eventually destroys the superconducting phase. 
Remarkably, in such systems, the quasi-particle energy gap vanishes more 
rapidly than the order parameter, admitting the existence of a gapless 
superconducting phase. The (mean-field) theory of the gap suppression was
explored in a classic work by Abrikosov and Gor'kov~\cite{AG}. Since this
pioneering work, it is now realized that this scheme applies more widely, 
encompassing general time-reversal ($\cT$) symmetry breaking perturbations 
(for a review see, e.g., Refs.~\cite{maki,Tinkham}). In the following,
we will consider the general phenomenology of a disordered superconducting
thin film subject to a homogeneous in-plane magnetic field.

\begin{figure}[h]
\begin{center}
        \epsfxsize=3in\epsfbox{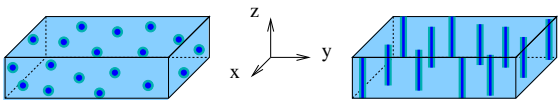}
\caption{\label{fig-dc}Schematic picture of a thin film with
    $\delta$-correlated disorder (left) and columnar defects (right).}
\end{center}
\end{figure}

Generally, a bulk superconductor subject to a weak magnetic field of 
magnitude $H<H_{c2}$ enters a Meissner phase~\cite{MeOc33}: the field is
expelled from the bulk, penetrating only a thin surface layer down to the 
London penetration depth. In the Meissner phase, the quasi-particle 
properties of the bulk states are not affected by the magnetic field and,
  thus, are largely insensitive to disorder. However,
if the dimensions of the superconductor are diminished to a scale comparable
to the penetration depth, field lines enter the sample and the 
superconductivity becomes strongly suppressed. Here the quasi-particle
properties deviate significantly from those of the unperturbed system: in
particular, the system can exhibit a gapless phase. At the level of 
mean-field, the properties of the thin film superconductor mirror those of 
the magnetic impurity system and are described by the
Abrikosov-Gor'kov (AG) 
phenomenology~\cite{maki}. However, beyond the level of mean-field, phase
coherence effects due to normal disorder strongly influence the long-range
properties of the system. In the following, we will investigate the 
properties of the gapless phase in the thin film superconducting system
focusing on two situations: in the first case, we will consider an 
arrangement of normal impurities which are drawn at random from a 
$\delta$-correlated distribution. In the second case, we will consider the
superconducting system subject to a disordered array of columnar
defects, i.e.~an impurity potential which does not depend on the coordinate 
perpendicular to the plane. The two arrangements are depicted schematically in 
Fig.~\ref{fig-dc}.

When applied to a superconductor, a magnetic field has two effects: firstly,
it induces a Zeeman splitting and, secondly, it couples to the orbital motion 
of the electrons. Both have the effect of suppressing superconductivity. 
Whereas, in a bulk system, the orbital effect usually dominates, in very 
thin films the opposite situation arises. The crossover can be estimated 
in the following way~\cite{KeAl98}: the critical magnetic field associated 
with the orbital effect is roughly determined by the condition that the flux 
threading an area spanned by the coherence length is of the order of one 
flux quantum, $H_{c2}\xi^2\simeq\phi_0$, where $\xi=\sqrt{D/(2\Delta)}$ is the coherence 
length in the dirty system with order parameter $\Delta$ and diffusion constant $D$. Now, if the
width of the film $d$ becomes smaller than the coherence length, $d\ll\xi$, this has to be 
replaced by the condition $H_{c2}^\parallel \xi d\simeq\phi_0$, where $H_{c2}^\parallel$
represents the in-plane field. I.e., the orbital critical field increases. 
The critical field $H_Z$ associated with the Zeeman splitting is independent 
of the width of the system. $H_Z$ is obtained from the condition that the 
energy splitting between up($\uparrow$)- and down($\downarrow$)-spins is 
roughly of the size of the order parameter, $g_L\mu_BH_Z\simeq\Delta$, 
where $g_L$ is the Land\'e $g$-factor and $\mu_B=e/(2m)$ the Bohr magneton. 
Comparing the two, 
one concludes that 
the orbital effect is dominant in suppressing superconductivity when
\begin{eqnarray*}
d>d_c\equiv g_L\,{\lambda_{\rm F}\xi\over \ell}, 
\end{eqnarray*}
where $\lambda_{\rm F}$ is the Fermi wavelength and $\ell$ the elastic 
mean free path.

In the following, we restrict attention to systems where $d>d_c$ and
the Zeeman splitting can be neglected. 
To be concrete, we will consider a thin film system where 
the relevant length scales are arranged in the following hierarchy:
\begin{eqnarray}
\lambda_{\rm F}\ll [d,\ell]\ll\xi.
\label{cond-SC}
\end{eqnarray}
The inequality $\lambda_{\rm F}\ll \ell$ defines the quasi-classical
regime while $\ell\ll\xi$ is the condition for the dirty limit. Finally, 
$\lambda_{\rm F}\ll d$ implies that the subband splitting due to size 
quantization is small and many subbands are occupied. Thus, from the 
point of view of the normal electrons, the system is effectively 
three-dimensional.

Before turning to the theoretical analysis, let us first briefly summarize 
the main results of this investigation. At the level of mean-field, the 
superconducting system with an in-plane magnetic field is described by the 
AG theory~\cite{AG}, independent of the nature of the normal disorder 
potential (diffusive or columnar). The parameter governing the suppression 
of the quasi-particle energy gap is set by 
\begin{eqnarray*}
\zeta\sim {D(Hd)^2\over \Delta}\sim \Phi_{d\xi}^2, 
\end{eqnarray*}
where $\Phi_{d\xi}$ is the flux through an area perpendicular to the field 
spanned by the width of the film $d$ and the superconducting coherence 
length $\xi$. As found for the magnetic impurity system~\cite{LaSi00}, 
within the gapped phase, the sharp edge in the quasi-particle density
of states (DoS) predicted 
by mean-field is softened by the nucleation of localized tail states which extend into the 
sub-gap region. In this sense, the mean-field gap edge becomes a mobility edge separating a region of bulk (weakly localized) quasi-particle states from strongly localized tail states. In the following, we will demonstrate that these sub-gap 
states are induced by optimal fluctuations of the random impurity potential 
and are reflected as inhomogeneous instanton field configurations of the 
mean-field equations. In the vicinity of the gap edge $E_{\rm gap}$, the 
energy scaling of these tails is universal~\cite{LaSi00}, depending only 
on the distance from $E_{\rm gap}$, the dimensionless parameter $\zeta$ 
and dimensionality. In $2d$, one obtains the result, 
\begin{eqnarray}
\nu(\epsilon<E_{\rm gap})\sim\exp\left[-a_2(\zeta)\nu_\Box D\,\frac{E_{\rm gap}
-\epsilon}\Delta\right],
\end{eqnarray}
non-perturbative in the inverse dimensionless conductance $1/(\nu_\Box
D)$, where $\nu_\Box$ is the DoS of the system when in the normal state.
Here $a_2$ is a known dimensionless function of the control parameter $\zeta$. 

At the mean-field level, the choice of disorder does not affect the results 
qualitatively. However, within the gapless phase, the low-energy physics 
depends sensitively on the character of the impurity potential and, in
particular, on whether or not it respects inversion symmetry. Within the 
diffusive regime $1/\tau>\epsilon>D/L^2$, the quasi-particle states 
exhibit a localized or ``spin-insulator'' phase with a localization
length $\xi_{\rm loc}$
which depends on the fundamental symmetry of the system~\cite{oppermann,Zirnbauer,Fisher}. However, it is
within the {\em ergodic} regime $\epsilon<{\rm max}[D/\xi_{\rm loc}^2,D/L^2]$,
where the influence of the inversion symmetry is most clearly exposed.
Here the quasi-particle properties become universal, dependent only on the
fundamental symmetry class~\cite{class,AlZi97}. A complete
classification scheme of ten symmetry classes~\cite{class},
corresponding to the ten large families of symmetric spaces identified
by Cartan, exists; four of these symmetry classes are relevant in the
context of superconductity~\cite{AlZi97}. Possible symmetries are 
time-reversal, spin-rotation, particle-hole and chiral symmetry. Now s-wave
superconductors possess particle-hole symmetry but, generically, do not exhibit 
chiral symmetry. Furthermore, they may or may not possess time-reversal 
and/or spin-rotation symmetry. Here, as one might expect, one finds that the diffusive film in a 
magnetic field belongs to `symmetry class C' (according to Cartan's notation) 
which corresponds to spin-rotation symmetry, but broken time-reversal 
symmetry. By contrast -- despite the presence of the 
magnetic field -- the film with columnar defects manifests a hidden symmetry
placing the superconductor in the higher symmetry class CI which 
is usually characteristic of systems possessing time-reversal invariance. 
The distinct symmetry classes manifest themselves in the energy
dependence of the density of states for $\epsilon\to0$. Thus, although 
superficially the suppression of superconductivity does
not respond to geometrical symmetries, manifestations of the
Berry-Robnik symmetry effect can be observed in the gapless phase. The 
distinct low-energy behavior predicted above is confirmed by numerics.

Finally, to complete our discussion, we turn to the consideration 
of the influence of an in-plane magnetic field on the spectral properties
of a normal metal-superconductor (NS) bi-layer. Here the interplay of
macroscopic quantum coherence phenomena in the superconductor and the
mesoscopic metal leads to interesting effects~\cite{FoFe01}. Generally, a superconductor brought into contact with
a normal system tends to impact aspects of its superconducting character
onto the latter. At the origin of this phenomenon is the Andreev
reflection~\cite{andreev} of electrons at the interface: an electron
with energy smaller than $\Delta$ may be retro-reflected off the
boundary as a hole, and a Cooper pair added to the condensate in the
superconductor. The coherent superposition of the incident electron with the
reflected hole leads to a non-vanishing pair amplitude within the
normal region. In particular, the density of states
within the normal region may develop a gap -- whose size depends on
the coupling $\gamma$ between the two systems. Here we consider the case of weak
coupling, where the induced gap $E_{\rm gap}^{(N)}=\gamma \;(\ll\Delta)$. 
A parallel magnetic field
counteracts this phenomenon by suppressing the energy gap even within
the superconductor. Below, we will show that the field necessary to suppress the induced gap in 
the normal region is much smaller than the field required to drive the 
superconductor into the gapless phase. However, the coupling to the
normal region also induces a non-vanishing DoS at low energies $E_{\rm gap}^{(N)}<\epsilon<\Delta$ in the 
superconductor.
Here we consider two different geometries, namely a planar NS bi-layer
as well as an NS cylinder with the magnetic field directed along the axis. In
the second system the enclosed flux leads to a periodic modulation of
the field effect as first observed in an experiment by Little and Parks~\cite{little-parks}. Only for small enough cylinders the system reaches the gapless phase in certain ranges of the magnetic field. By further decreasing the radius, superconductivity is completely suppressed around half-integer flux quanta threading the cylinder, as predicted by de Gennes~\cite{deGennes} and very recently verified experimentally by Liu {\it et al.}~\cite{newexp}.

The paper is organized as follows:
in Sec.~\ref{sec-filmdiff}, we first study the suppression of
superconductivity due to the parallel field at the mean-field
level. In Sec.~\ref{sub-tails} we show how the results for the sub-gap 
tail states, recently established for the magnetic
impurity system, translate to thin films in parallel fields.
Subsequently, in Secs.~\ref{sec-filmcol} and \ref{sec-C_1}, we search
for manifestations of the Berry-Robnik symmetry effect on the properties of the superconducting phase. In Sec.~\ref{sec-NS}, we discuss the properties of thin disordered NS bi-layers. Finally, we conclude in Sec.~\ref{sec-conclusions}.

\section{Diffusive film}\label{sec-filmdiff}

Before turning to the columnar defect system, let us begin with a 
discussion of the most generic case: a thin film with just a `normal'
$\delta$-correlated white noise disorder potential. Here we are interested
in the limit, where -- in addition to the conditions~(\ref{cond-SC})
specified above --
$\ell\ll d$ which implies
diffusive motion in all three directions.

Building on the field theory approach to the study of weakly disordered 
systems~\cite{We79,EfLa80,Finkel83} (for a review see, e.g., Ref.~\cite{efetov}), the 
extension to the consideration of disordered superconducting systems 
follows straightforwardly~\cite{Oppermann,SIT,AST}. Therefore, here we 
will only
briefly summarize the main elements in the construction of the 
field theory of the disordered superconductor in the framework of the 
non-linear $\sigma$-model (NL$\sigma$M). Using this formulation, we will thereafter 
investigate the response of the superconducting film to an in-plane
magnetic field.

The superconducting system is described by the Gor'kov Hamiltonian
\begin{eqnarray*}
{\cal H}=\left(\begin{matrix}{{\cal H}_0 &\Delta\\\Delta^*& 
-{\cal H}_0^T}\end{matrix}\hspace*{-1.1cm}\right)_{\sc ph}
\end{eqnarray*}
with ${\cal H}_0={\bf p}^2/(2m)-\epsilon_{\rm F}+V({\bf r})$. Here, the matrix structure refers to the particle/hole ($\sc ph$) space, 
$\epsilon_{\rm F}$ denotes the Fermi energy, and $V({\bf r})$ represents 
the quenched random impurity potential whose distribution is characterized 
by the mean scattering time~$\tau$. We consider the 
potential to be drawn from a Gaussian white noise distribution, 
$\langle V({\bf r})V({\bf r}')\rangle_V=(2\pi\nu_0\tau)^{-1}\delta^{(3)}({\bf r}
-{\bf r}')$, where $\nu_0$ is the average DoS of the normal system.

The starting point of the analysis is the functional field integral
for the generating functional~\cite{Oppermann,SIT,AST,Zirnbauer}
\begin{eqnarray*}
{\cal Z}[j]=\int D[\Psi,\bar\Psi] \;e^{-S[\bar\Psi,\Psi]+\int d\br\,(\bar\Psi j+\bar j\Psi)},
\end{eqnarray*} 
where $\Psi$ represent 8-component supervector fields, incorporating a boson/fermion space as 
well as the particle/hole and a charge conjugation space: to properly account for all channels of quantum interference, it is standard practice to affect a doubling of the field space to accommodate the particle-hole symmetry $\cH=-\sph_2\cH^T\sph_2$, where $\sigma^{\sc ph}_2$ is a Pauli matrix in particle/hole space. This additional space is referred to as the charge conjugation sector. The
introduction of commuting (bosonic) as well as
anti-commuting (fermionic) fields ensures the normalization of the generating
functional in the absence of sources, $\cZ[0]=1$. Following
Ref.~\cite{Zirnbauer}, the superfields are not independent, but obey
the condition $\bar\Psi=(i\sph_2\!\otimes\!\eta \,\Psi)^T$, where
$\eta=\scc_1\otimes E_{\sc{bb}}-i\scc_2\otimes E_{\sc{ff}}$ (and
$E_{\sc{bb}}={\rm diag\,}(1,0)_{\sc bf}$, $E_{\sc{ff}}={\rm
  diag\,}(0,1)_{\sc bf}$ are projectors onto the boson-boson and
fermion-fermion block, respectively). Taking our notation from Ref.~\cite{Zirnbauer}, the action 
assumes the canonical form
\begin{eqnarray*}
&&S[\bar\Psi,\Psi]=i\int d{\bf r}\; \bar\Psi(\epsilon^-\sigma^{\sc cc}_3-
{\cal H})\Psi,
\end{eqnarray*}
where $\sigma^{\sc cc}_3$ is a Pauli matrix in charge conjugation
space, $\epsilon^-=\epsilon-i0$, and ${\cal H}$ represents the Gor'kov
Hamiltonian introduced above.

To explore the low-energy properties of the superconducting system,
after ensemble averaging over the random impurity
distribution, the functional integral over the supervector fields $\Psi$ 
can be traded for an integral involving supermatrix fields $Q$. Physically, 
the fields $Q$, which vary slowly on the scale of the mean-free 
path $\ell$, describe the soft modes of density relaxation --- the 
diffusion modes. In the quasi-classical limit ($\epsilon_{\rm F}\gg 
1/\tau$), the action for $Q$ is dominated by the saddle-point field 
configuration. In the dirty limit ($1/\tau\gg\Delta$), the
saddle-point equation admits
the solution $Q_0=\sigma^{\sc ph}_3\otimes\sigma^{\sc cc}_3$.
However, in the limit $\epsilon\to 0$,
the saddle-point is not unique but spans an entire manifold parameterized
by the unitary transformations $Q=TQ_0T^{-1}$. Taking into account
slow spatial and temporal fluctuations of the fields, the low-energy
long-range properties of the weakly disordered superconducting system
are described by an action of non-linear $\sigma$-model 
type~\cite{efetov,Oppermann,SIT,AST,Zirnbauer}
\breakon
\begin{eqnarray}
S[Q]=-\frac{\pi\nu_0}{8}\int d{\bf r}\; {\rm Str}\,\left[D(\partial
Q)^2-4(i\epsilon^-\sigma_3^{\sc ph}\otimes\sigma_3^{\sc cc}-\Delta
\sigma_2^{\sc ph})Q\right],
\label{S-gen}
\end{eqnarray}
\breakoff
\noindent
where the supermatrix field obeys the non-linear constraint $Q^2=\openone$. 
Furthermore, the matrices~$Q$ obey the the symmetry relation 
$Q=\sph_1\otimes\eta\, Q^T(\sph_1\otimes\eta)^T$,
reflecting the symmetry properties of the dyadic 
product $\Psi\otimes\bar\Psi\sph_3$. 
Finally,
due to gauge invariance properties of the action, the incorporation of a 
magnetic field amounts to replacing the derivatives in Eq.~(\ref{S-gen}) by
their covariant counterpart $\tilde\partial=\partial
-i\bA[\sph_3,\,.\,]$. This completes the formulation of the disordered superconducting system
as a functional field integral involving the supersymmetric NL$\sigma$M action.
Our interest here is in the thermodynamic DoS obtained as 
\begin{eqnarray*}
\nu(\epsilon) = \pi^{-1} \int  {d{\bf r}\over V} \,\Im 
\left[G({\bf r},{\bf r};\epsilon^-)\right]\,.
\end{eqnarray*}
Making use of the generating functional, it is straightforward to show that the impurity
averaged DoS can be obtained from the identity
\begin{eqnarray}
\left\langle\nu(\epsilon)\right\rangle={\nu_0\over 8}\int {d{\bf r}\over 
V}\, \Re \big\langle {\rm Str}\left[
\sigma_3^{\sc bf}\otimes\sigma_3^{\sc ph}\otimes\sigma_3^{\sc cc} \,Q \right]
\big\rangle_{Q}\,,
\label{gendos}
\end{eqnarray}
where $\langle\cdots\rangle_{Q}=\int DQ \,\cdots\, 
e^{-S[Q]}$.

Thus, the starting point of our analysis is the conventional
NL$\sigma$M, Eq.~(\ref{S-gen}), for
a three-dimensional superconducting system subject to a magnetic field~\cite{efetov,AST},
where the vector potential reads $\bA=-Hz{\bf e}_y$.
The typical scale of variation of the $Q$-fields is set by the
coherence length $\xi$. Therefore, since $d\ll\xi$, the matrices $Q$ are constant along the
$z$-direction, allowing integration along $z$ to be performed explicitly. Making use of the identities 
\begin{eqnarray}
\frac1d\int\limits_{-d/2}^{d/2} dz \, \bA=0, \qquad \frac1d\int\limits_{-d/2}^{d/2} dz \, \bA^2=\frac1{12}(Hd)^2,\nn
\end{eqnarray}
one finds that
\begin{eqnarray}
S=-\frac{\pi\nu_\Box}{8}\int \!d^2r\, \Tr\Big[D(\partial
  Q)^2-\frac\kappa2[\sph_3,Q]^2-&&\nn\\
-4(i\epsilon^-\sph_3\!\otimes\!\scc_3\!-\!\Delta\sph_2)Q\Big],&&\label{S-kappa}
\end{eqnarray}
where $\kappa=D(Hd)^2/6$ and $\nu_\Box$ is the DoS of the
two-dimensional system, i.e.~here $\nu_\Box=\nu_0 d$. In this planar geometry, the paramagnetic term, which is crucial for the Meissner effect, vanishes from the action.

Note that the choice of gauge is important here: the physical gauge to
choose is the London gauge, $\nabla\cdot\bA=0$ and $A_z(\pm \,d/2)=0$. Both
conditions are fulfilled by $\bA=-Hz{\bf e}_y$. The above requirements 
originate form the fact that, in a superconductor,
the vector potential is associated with a supercurrent
$\bj_s=n_s\bA/m$, where $n_s$ is the density of Cooper pairs. The 
first condition tells us that no net current is generated while the second condition does not
allow a supercurrent to flow through the superconductor-vacuum
boundary.
Thus, when integrating along $z$, we have fixed the gauge, i.e.~the resulting
action is not gauge invariant. Therefore, the magnetic field does not
appear within a covariant derivative, but as an additional diamagnetic
term~$\sim \kappa[\sph_3,Q]^2$. This distinguishes the ``thick'' film, $d\gg\lambda_{\rm F}$,
from the single-channel case (i.e., the truly two-dimensional system), where the magnetic field can be gauged
out and, thus, has no influence.

\subsection{Mean-field analysis}\label{sub-diffMF}

\begin{figure}[h]
\begin{center}
\epsfxsize=3in\epsfbox{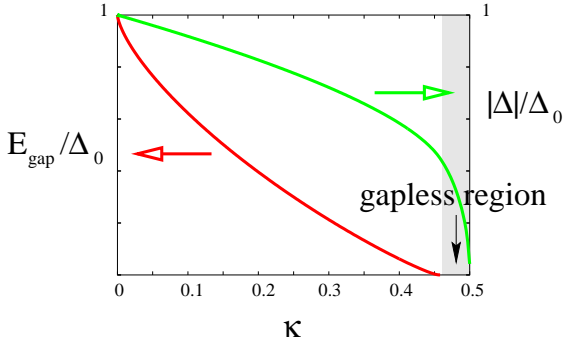}
\caption{\label{fig-gapphase} Dependence of the energy gap $E_{\rm gap}$ and the
order parameter $\Delta$ on $\kappa$. Gapless sueprconductivity occurs 
in the narrow range $\Delta_0\,\exp[-\pi/4]\leq\kappa\leq\Delta_0/2$.}
\end{center}
\end{figure}

As usual, at the mean-field level, the density of states is obtained by subjecting the 
action~(\ref{S-kappa}) to a saddle point analysis. Varying the action 
with respect to $Q$, subject to the non-linear constraint $Q^2=\openone$,
one obtains the saddle-point equation
\begin{eqnarray}
D\partial(Q\partial Q)-\frac\kappa2[\sph_3 Q\sph_3,Q]-&&\nn\\
-[i\epsilon^-\sph_3\otimes\scc_3-\Delta\sph_2,Q]\eq 0.
\end{eqnarray}
The latter can be identified as the Usadel equation~\cite{usadel} for the
average quasi-classical Green function, supplemented by an additional term 
due to the parallel magnetic field. With the {\em Ansatz} $Q=\cosh\hat
\theta\,\sph_3\otimes\scc_3+i\sinh\hat\theta\,\sph_2$, where 
$\hat\theta=\theta_{\sc b}E_{\sc bb}+i\theta_{\sc f}E_{\sc ff}$, one obtains
\begin{eqnarray}
D\partial^2\hat\theta-2i(\epsilon\sinh\hat\theta-\Delta\cosh\hat\theta)
-\kappa\sinh(2\hat\theta)=0.
\label{sp-B}
\end{eqnarray}
Taking $\theta$ to be homogeneous in space, and defining
\begin{eqnarray}
\tilde \epsilon=\epsilon-\frac i2\kappa\cosh\hat\theta, \qquad \tilde
\Delta=\Delta+\frac i2\kappa\sinh\hat\theta,\nn
\end{eqnarray}
the equations for $\tilde\epsilon$ and $\tilde\Delta$ take the BCS form 
and, thus, admit the diagonal solution $\tilde\epsilon/\tilde\Delta=\coth\theta$. Then, in 
terms of the `bare' parameters $\epsilon$ and $\Delta$, the saddle-point 
equation~(\ref{sp-B}) can be brought to the conventional AG form~\cite{AG} 
\begin{eqnarray}
\frac\epsilon\Delta=u\left(1-\zeta\frac1{\sqrt{1-u^2}}\right),
\label{ag-n}
\end{eqnarray}
where $u\equiv\coth\theta$. Following for example Ref.~\cite{maki}, by extremizing this equation and defining $\zeta=\kappa/\Delta$, one obtains
the quasi-particle energy gap
\begin{eqnarray*}
E_{\rm gap}=\Delta(1-\zeta^{2/3})^{3/2}.
\end{eqnarray*}
When combined with the self-consistency equation for the order parameter~\cite{maki}
\begin{eqnarray*}
\Delta=\frac{\pi g\nu_\Box}\beta\sum_n\frac1{\sqrt{1+u_n}}\,,
\end{eqnarray*}
where we have shifted to Matsubara frequencies,
$i\epsilon^-\to\epsilon_n=(2n+1)\pi/\beta$, one finds that, at $T=0$,
superconductivity is destroyed when $\kappa_\Delta=\Delta_0/2$, where
$\Delta_0$ is the order parameter in the absence of a magnetic
field. The gapless phase sets in at the smaller value $\kappa_{\rm
  g}=\Delta_0\exp[-\pi/4]$. In the following, we will be able to
determine the order parameter self-consistently, and $\Delta$ will be
understood as the renormalized order parameter, even if not stated
explicitly. The suppression of the energy gap $E_{\rm gap}$ and the
order parameter $\Delta$ are shown in Fig.~\ref{fig-gapphase}.

Finally, in the mean-field approximation, the DoS takes the form
\begin{eqnarray}
\nu(\epsilon)\eq\frac{\nu_\Box}8\int\frac{d^2r}V\,\Re\Tr[\sbf_3\otimes
\sph_3\otimes\scc_3\, Q(\br)]\\
\eq\nu_\Box\,\Re[\cosh\theta].\nn
\end{eqnarray}
Inserting the solution determined by Eq.~(\ref{ag-n}), this yields the 
characteristic AG DoS, exhibiting a square-root edge in the gapped phase.

\subsection{Sub-gap states}\label{sub-tails}

According to the mean-field AG theory, the superconducting system is 
expected to exhibit a gapped phase over a wide range of parameters with 
the DoS varying as a square-root of the energy difference
$\epsilon-E_{\rm gap}$ in the vicinity 
of the gap edge $E_{\rm gap}$. Recently, studies in the related context of 
a superconducting system perturbed by magnetic impurities have shown that 
the hard edge predicted by the AG theory is
untenable~\cite{BaTr97,LaSi00}.
In particular, in the diffusive system, it has been shown that optimal 
fluctuations of the random impurity potential lead to the nucleation of 
localized sub-gap quasi-particle states which soften the gap 
edge~\cite{LaSi00}. 

Within the framework of the supersymmetric field theory, the precise 
mechanism of sub-gap state formation has been elucidated in 
Ref.~\cite{LaSi00}. There, it was shown that, in addition to the 
homogeneous AG solution, the saddle-point 
equation~(\ref{sp-B}) accommodates a degenerate manifold of spatially 
inhomogeneous instanton or bounce solutions. Referring to
Ref.~\cite{LaSi00} 
for details, to exponential accuracy, the sub-gap DoS takes the form
\begin{eqnarray}
\nu(\epsilon)\sim \exp\left[-c_2\nu_\Box D
\,\zeta^{-2/3}(1-\zeta^{2/3})^{-1/2}\,
\frac{E_{\rm gap}\!-\!
\epsilon}{\Delta}\right],
\label{nu-exp}
\end{eqnarray}
where $c_2$ is a numerical constant. Within the same theory, one finds 
the sub-gap states are confined to droplets of size 
\begin{eqnarray*}
r_{\rm drop}(\epsilon)=6^{1/4}\,\xi\left(\frac{\Delta}{E_{\rm
      gap}}\right)^{1/12}\left(\frac{E_{\rm gap}-\epsilon}\Delta\right)^{-1/4},\nn
\end{eqnarray*}
diverging upon approaching the gap edge. The mean-field gap edge
$E_{\rm gap}$ therefore assumes the significance of a mobility edge separating 
localized sub-gap states from extended bulk states.

Being confined to a region of size $r_{\rm drop}\gtrsim \xi\gg \ell\gg 
\lambda_{\rm F}$, it is evident that the physical mechanism of sub-gap state 
formation is quasi-classical in origin, relying on optimal fluctuations 
of the random impurity potential. Moreover, in contrast to Lifshitz 
semiconductor band-tail states~\cite{Lif65}, each droplet leads to the 
nucleation of an entire band of localized states. Qualitatively, the 
physical mechanism of sub-gap state formation is connected to mesoscopic 
fluctuations in the phase sensitivity of the electron wavefunction. In 
regions where the sensitivity is high, the impact of a time-reversal 
symmetry breaking such as that imposed by the magnetic impurities or 
parallel magnetic field is stronger, and the effective scattering rate 
$\zeta$ is enhanced.

Note that, instead of applying a magnetic field, one might consider driving a
supercurrent through the system. As emphasized in Ref.~\cite{maki}, this 
leads to the same AG mean-field results. However, in this case, due to the presence of a paramagnetic term in the action, the mean-field gap edge is robust, i.e.~no sub-gap states are generated. 

\section{Columnar defects}\label{sec-filmcol}

In the normal system, previous studies~\cite{MAA01} have shown that the 
properties of a thin film in an in-plane magnetic field depend sensitively 
on the nature of the impurity potential. Do the quasi-particle properties
of the disordered superconductor exhibit a similar sensitivity?

In the normal system, $z$-inversion symmetry effectively compensates for 
the time-reversal symmetry breaking of the magnetic field, driving the 
system into the orthogonal symmetry class~\cite{BeRo86}. As such, one might expect 
that the inversion symmetry restores the validity of the Anderson theorem, 
rendering the superconducting phase insensitive to the disorder potential.
In fact, constraints imposed by self-consistency of the order parameter 
prevents the existence of a mechanism that could cancel the effect of the
magnetic field. The reason is that here we are dealing with an interacting 
problem: with the formation of Cooper pairs, it is not possible to replace 
time-reversal by any other symmetry. Indeed, at the level of the mean-field
AG theory, the influence of columnar defects is indistinguishable from 
that of the diffusive scatterers -- only the parameter $\zeta$ is modified. 

However, as we will see in Sec.~\ref{sec-C_1}, the Berry-Robnik phenomenon 
described in Ref.~\cite{BeRo86} is not completely ineffective in the 
superconducting phase. Taking into account fluctuations in the gapless 
phase, we will show that, while the diffusive film belongs to the 
fundamental symmetry class C (corresponding to a disordered superconductor
in a magnetic field), with columnar defects, the system belongs to the higher
symmetry class CI (characteristic of the time-reversal invariant superconductor). 
The result is a substantial modification of the low-energy behavior: in 
the diffusive film, quantum interference phenomena in the particle/hole
channel induce a microgap structure with a DoS varying as $\nu(\epsilon)
\sim \epsilon^2$, while in the film with columnar defects 
$\nu(\epsilon)\sim \epsilon$.

To be specific, let us consider a model of a thin film superconductor 
subject to a random (impurity) potential which varies only along the in-plane
directions. In the absence of a magnetic field or superconducting order 
parameter, the quasi-particle Hamiltonian can be subdivided into different 
subbands labeled by an index $k$. The spectral properties of each subband are 
described by a two-dimensional NL$\sigma$M action of conventional type. 
The derivation of an effective low-energy action follows closely the 
normal case in Ref.~\cite{MAA01}. The Gor'kov Hamiltonian of the system 
now reads
\begin{eqnarray}
{\cal H} = \Big(-\frac{\tilde\partial^2}{2m}+W(z)-V(x,y)\Big)\sph_3+
\Delta(z)\sph_2,
\end{eqnarray}
where $\tilde\partial=\partial+iHz{\bf e}_y\sph_3$, $W$ is the confining 
potential, and $V$ represents an impurity potential drawn at random
from the white-noise $\delta$-correlated distribution with zero mean, and
correlation $\langle V(\br)V(\br')\rangle=(2\pi\nu_\Box\tau)^{-1}
\delta^{(2)}(\br-\br')$, where $\br^{(\prime)}$ are `in-plane' 
two-component vectors. 
 
Diagonalizing the $z$-dependent part of the problem, and representing
$\cH$ in the basis of the (real) eigenfunctions $\{\phi_k\}$, i.e.~$\cH_{kk'}
=\int dz\,\phi_k\,\cH\,\phi_{k'}$ where $(-\partial_z^2/(2m)+W(z)-\epsilon_k)\phi_k=0$, the vector potential $\bA=-Hz{\bf e}_y$ 
as well as the order parameter become matrices in $k$-space: 
\begin{eqnarray}
\bA_{kk'}\eq-H{\bf e}_y\int dz\,\phi_k(z)z\phi_{k'}(z),\nn\\
 \Delta_{kk'}\eq\int dz \,\phi_k(z)\Delta(z)\phi_{k'}(z).\nn
\end{eqnarray}
Let us emphasize that, if the system possesses inversion symmetry 
$z \to -z$, the matrix element $A_{kk'}$ differs from zero only if $k+k'$ 
is odd; in particular, $A_{kk}=0$. For simplicity, here we only consider the
fully symmetric case~\cite{fn1}.

Under the further assumption that the subband spacing $|\epsilon_k-
\epsilon_{k'}|$ is larger than the scattering rate~\cite{fn0}, one finds that only 
the diagonal components of the order parameter are non-vanishing. 
Starting from the conventional superconducting 2$d$ NL$\sigma$M action for 
the $k$ subbands and turning on an in-plane magnetic field, it is 
straightforward to show that the total effective action assumes the form
\breakon
\begin{eqnarray}
S\eq-\frac{\pi\nu_\Box}8\int d^2r \sum_k \Tr\left[D_k (\partial Q_k)^2
-4\left(i\epsilon^-\sph_3\otimes\scc_3-\Delta_{kk}\sph_2\right)Q_k
-2\sum_{kk'} \cX_{kk'} \,\sph_3Q_k\sph_3Q_{k'}\right],
\label{S_k}
\end{eqnarray}
\breakoff
\noindent
where $\cX_{kk'}=D_{kk'}A_{kk'}A_{k'k}/(1+(E_{kk'}\tau)^2)$. Furthermore,
$D_{kk'}=(D_k+D_{k'})/2$ (with $D_k$ denoting the diffusion constant of subband
$k$) and $E_{kk'}=\epsilon_k-\epsilon_{k'}$. 
Crucially, from this result we see that there exists no linear coupling
of $Q$ to the vector potential -- a paramagnetic term does not appear. 

To proceed, as before we subject the action (\ref{S_k}) to a mean-field analysis. Varying
the action with respect to fluctuations of $Q_k$, one obtains the
modified (set of coupled)
Usadel equations
\begin{eqnarray}
D_k\partial\left(Q_k\partial Q_k\right)-\left[i\epsilon^-\sph_3\otimes\scc_3-\Delta_{kk}\sph_2,Q_k\right]-&&\nn\\
-\sum_{k^\prime} \cX_{kk^\prime}
\left[\sph_3 Q_k \sph_3,Q_{k^\prime}\right]\eq0.\nn
\end{eqnarray}
Applying the Ansatz 
$Q_k=\cosh\hat\theta_k\sph_3\otimes\scc_3+i\sinh\hat\theta_k\sph_2$
with $\theta_k$ homogeneous in the in-plane coordinates, the mean-field 
equation assumes the form
\begin{eqnarray}
\epsilon^-\sinh\theta_k\!-\!\Delta_{kk}\cosh\theta_k\!-\!i\sum_{k^\prime}\cX_{kk^\prime}
\sinh\left(\theta_k\!+\!\theta_{k^\prime}\right)=0.\label{sp}
\end{eqnarray}
In principle, this equation has to be solved in parallel with the 
self-consistent equation for the order parameter
\begin{eqnarray}
\Delta_{kk'}=\frac{\pi g\nu_\Box}\beta \sum_n\sin\theta_{k,n}\,\delta_{kk'},\label{sc}
\end{eqnarray}
where $g$ is the effective BCS coupling constant and $n$ is a Matsubara index.

Analyzing the saddle point equation~(\ref{sp}), it can be easily seen that the field
dependent term $\sum_{k'}\cX_{kk'}\sin(\theta_k+\theta_{k'})$ vanishes
if we choose the solution $\theta_k=(-1)^k\theta$ (i.e.~since $\cX_{kk'}=0$ 
for $k+k'$ even). Thus, there seems
to be one
mode which is not affected by the magnetic field. However, this would
imply that the order parameter, too, must have an alternating sign,
i.e.~$\Delta_{kk}=(-1)^k\Delta$. Recalling the definition
$\Delta_{kk}=\int dz\, \Delta(z)\phi_k^2$, this is not feasible. Thus, the solution above is ruled
out~\cite{fn2} and, therefore, on the mean-field level, the symmetry
mechanism is ineffective.

A more natural choice seems to be a spatially homogeneous order parameter.
Unfortunately, for a general model, the solution of the coupled Eqs.~(\ref{sp})
and (\ref{sc}) does
not seem to be readily accessible analytically. However, to gain some insight into the
nature of the general solution, we will specialize further consideration
to the particular case in which only the lowest two subbands are coupled.

With $\cX_{12}=\cX_{21}\equiv \cX$ the equations for $\theta_1$ and $\theta_2$
coincide. Therefore, setting $\theta\equiv\theta_1=\theta_2$, which
implies that $\Delta_{11}=\Delta_{22}\equiv\Delta$, the mean-field equation takes
the form reminiscent of the AG equation,
\begin{eqnarray}
\epsilon\sinh\theta-\Delta\cosh\theta-i\cX\sinh\left(2\theta\right)=0.\nn
\end{eqnarray}
As with the diffusive film, the application of a 
strong in-plane field suppresses the order parameter and allows for
the existence of a gapless phase. According to the AG theory, the 
superconductor enters the gapless phase when $\zeta\equiv 2\cX/\Delta\simeq 1$.

If $E_{12}\tau\ll1$, the parameter $\zeta$ is of the same form as that found in
the diffusive case, i.e.~$\zeta\sim D(Hd)^2/\Delta$. In the opposite
limit, $\zeta$ is greatly reduced because the
wide subband spacing restricts the motion in $z$-direction. Now,
$\zeta\sim D(Hd)^2/((E_{12}\tau)^2\Delta)$, and, thus, higher magnetic
fields have to be 
applied in order to reach the gapless phase. As in the diffusive case, 
the hard edge in the gapped phase is compromised due to fluctuations --
see the discussion above -- and exponentially small tails in the
sub-gap region arise.

More generally, for many subbands, one would expect the same qualitative
picture to hold -- although $\Delta_{kk}$ might slowly depend on $k$.

\begin{figure}[h]
\begin{center}
\epsfxsize=2.5in\epsfbox{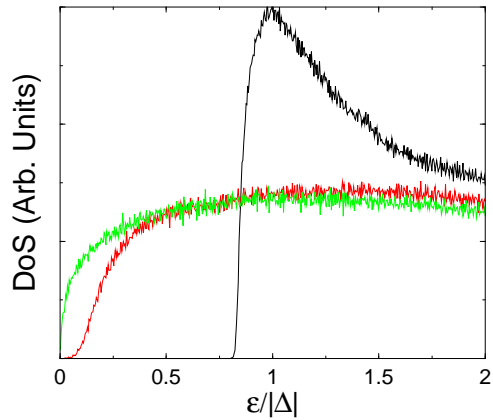}
\caption{\label{fig-numerics}Numerical results: DoS.
    Upon increasing the magnetic field, the energy gap closes and the
    BCS singularity disappears.}
\end{center}
\end{figure}

\begin{figure}[h]
\begin{center}
\epsfxsize=2.5in\epsfbox{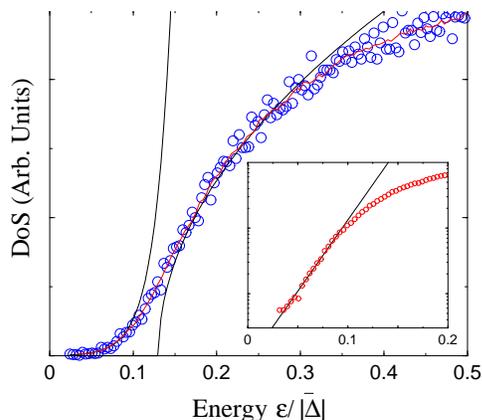}
\caption{\label{fig-num_le}Numerical results: sub-gap
    DoS. The solid lines show the square-root edge and the
    exponentially small tails. In the inset, the same data are plotted on a linear-log scale.}
\end{center}
\end{figure}

The effect of gap suppression is born out in a simple numerical simulation. 
Fig.~\ref{fig-numerics} shows the quasi-particle DoS for a two 
subband tight-binding model with $2\times20\times20$ sites when subject to an
in-plane magnetic field. The energy is measured in units of the (unperturbed) 
order parameter. The three curves correspond to different values of
the magnetic field. Details of the result at intermediate fields are
magnified in Fig.~\ref{fig-num_le}. The mean-field square-root edge as
well as the exponentially small tails are indicated. Furthermore, the inset shows
the linear energy dependence of the sub-gap DoS exponent,
c.f.~Eq.~(\ref{nu-exp}), on a linear-log scale: $\ln\nu(\epsilon<E_{\rm
  gap})\sim E_{\rm gap}-\epsilon$.

\section[Gapless phase]{Phase coherence properties of the gapless
  phase: massless fluctuations and the soft mode action}\label{sec-C_1}

While, at the level of mean-field, all perturbations (i.e.~magnetic
impurities as well as parallel fields in films with different disorder 
potentials) follow the same AG phenomenology,
it is interesting to note that differences show up in the spectrum of soft fluctuations. The latter are responsible for the long-range spectral and localization properties of the quasi-particles in the gapless phase. In contrast to the magnetic impurity
model~\cite{AG,LaSi00} which belongs to symmetry class D (due to
broken time-reversal and {\em spin-rotation} symmetry), here
the soft fluctuations around the mean-field should be described by an
effective action belonging to symmetry class C 
(broken $\cT$-invariance)~\cite{AlZi97}. Therefore, according to the considerations of Ref.~\cite{Zirnbauer,Fisher}, one expects localization of the
quasi-particle  
states in the gapless phase. In fact, below we will see that the fluctuations are sensitive
to the very nature of the impurity scattering. It turns out that the film with columnar defects is described by the
higher symmetry class CI, implying a modified localization length~\cite{oppermann,Fisher}.

To assess the low-energy properties of the system, we have to first identify 
the soft modes of the action.
For frequencies $\epsilon\to0$, the saddle point is not unique, but
spans a degenerate manifold $Q=TQ_{\sc{sp}}T^{-1}$ with
$T=\exp[W]$ and $\{Q_{\sc{sp}},W\}=0$. The symmetries of the system impose certain conditions on 
the generators $W$.

\subsection{Diffusive film}\label{sub-fdiff}

The choice of generators $W$ is dictated by the presence of the order
parameter and the magnetic field. This leads to the following conditions:
\begin{itemize}
\item Firstly, $W$ has to commute with the order parameter,
\[ [\sph_2,W]=0. \]
\item Secondly, since time-reversal symmetry is broken by the magnetic field, $W$
  has to fulfill the further restriction
\[ [\sph_3,W]=0. \]
\end{itemize}
Thus, $W=\openone^{\sc{ph}}\otimes W_s$. Taken together, these restrictions limit field fluctuations to those belonging to symmetry class C which describes superconducting
systems with spin-rotation symmetry, but broken $\cT$-invariance. The
corresponding integration manifold of class C is Osp(2$|$2)/Gl(1$|$1).
Expanded in the generators,
the soft mode action reads
\begin{eqnarray}
S_{Q_s}=-\frac{\pi\nu_\Box}{4}\int d^2r\,\Tr\big[D\cosh^2\theta(\partial Q_s)^2+&&\nn\\
+4i\epsilon\cosh\theta\scc_3Q_s\big],&&
\end{eqnarray}
where $Q_s=T_s\scc_3T_s^{-1}$ and $T_s=\exp[W_s]$.

On energy scales $\epsilon<E_c=D|\cosh\theta|/L^2$, the system enters the universal zero-dimensional regime. Here the properties of the action are dominated by the zero spatial mode and lead to~\cite{AlZi97}
\begin{eqnarray}
\nu(\epsilon)=\nu(E_c)\left(1-\frac{\sin(2\pi\epsilon/\delta)}{2\pi\epsilon/\delta}\right),
\end{eqnarray}
where $\delta=1/(\nu(E_c)L^2)$. I.e.~for $\epsilon\to0$, the DoS vanishes quadratically,
\begin{eqnarray}
\frac{\nu(\epsilon)}{\nu(E_c)}\simeq\frac23\pi^2\left(\frac\epsilon\delta\right)^2.\nn
\end{eqnarray}
This is to be contrasted with the low-energy behavior of the DoS in
the case of columnar defects, where the system possesses
the $\cP_z$-symmetry.

\subsection{Columnar defects}\label{sub-fcol}

Here instead of a single generator $W$, one has to consider a set of generators $W_k$.
\begin{itemize}
\item Once again, $W_k$ has to commute with the order parameter,
\[ [\sph_2,W_k]=0. \]
\item However, even though time-reversal symmetry is broken by the magnetic
  field, the generators do not have to obey $[\sph_3,W_k]=0$. Due to
  $\cP_z$-symmetry, which causes all elements ${\cal X}_{kk'}$ with
  $k+k'$ even to vanish, it is sufficient to require
\[ W_{k'}=\sph_3W_k\sph_3 \qquad {\rm for} \enspace k+k' \enspace {\rm odd}.\] 
I.e.~one generator, take e.g.~$W_0$, can be chosen
`freely'. Then, the others are determined through the condition
\[W_k=(\sph_3)^kW_0(\sph_3)^k,\] 
or: $W_k=W_0$ if $k\in2{\mathbb{N}}$, and
$W_k=\sph_3W_0\sph_3$ if $k\in2{\mathbb{N}}+1$. 
\end{itemize}
Thus, the second condition here only imposes certain relations between 
different $W_k$, but does not restrict the structure of $W_k$ in
particle-hole space. This corresponds to the higher symmetry class
CI. Now the
integration belongs to the group manifold Osp(2$|$2). Again we find a manifestation of the Berry-Robnik symmetry
phenomenon: the low-energy properties of the gapless phase are determined 
by the symmetry class associated with systems possessing time-reversal invariance.

Taking into account these fluctuations, the corresponding soft mode action reads
\begin{eqnarray}
S_{Q_s}=-\frac{\pi\nu_\Box}{8}\int d^2r\,\Tr\Big[D_k\cosh^2\theta_k(\partial Q_s)^2+&&\nn\\
+4i\epsilon\cosh\theta_k\sph_3\otimes\scc_3Q_s\Big],&&
\end{eqnarray}
where $Q_s=T_s\sph_3\otimes\scc_3T_s^{-1}$. Here $T_s=\exp[W_0]$, and $W_0$ fulfills
the conditions specified above. Once again,
properties of the class CI are available in the literature~\cite{AlZi97}. In particular, for small energies, one
obtains
\begin{eqnarray}
\frac{\nu(\epsilon)}{\nu(E_c)}\eq\frac\pi2\int\limits_0^{\pi\epsilon/\delta}\frac{dz}zJ_0(z)J_1(z)=\frac{\pi^2}4\epsilon/\delta+{\cal O}(\epsilon^3),
\end{eqnarray} 
showing the DoS to vanish linearly as $\epsilon\to0$.

\begin{figure}[h]
\begin{center}
\epsfxsize=2.5in\epsfbox{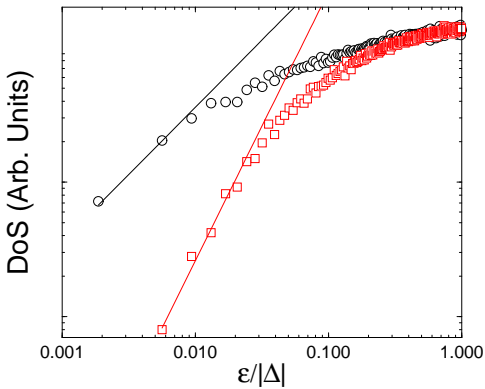}
\caption{\label{fig-num_dc}Numerical results: log-log
    plot of the low-energy DoS in the gapless phase for a diffusive film (open squares) and a film
    with columnar defects (open circles).}
\end{center}
\end{figure}

The predicted low-energy behavior can be verified numerically. In Fig.~\ref{fig-num_dc},
the density of states at low energies is compared for the two
cases. On the log-log scale one can read off the exponent $\alpha$
governing the energy dependence, $|\epsilon|^\alpha$. At low energies, 
the two lines
with slope $\alpha_{\rm C}=2$ and $\alpha_{\rm CI}=1$ --
characteristic of the symmetry classes C and CI -- fit the data for the diffusive film and the
film with columnar defects, respectively.

Having studied the influence of a parallel magnetic field on the properties of the superconducting film,
we now turn our attention to the mean-field properties of 
normal metal-superconductor hybrid systems.

\section{NS hybrid systems}\label{sec-NS}

The properties of thin disordered NS bi-layers have been studied in a recent work by Fominov and Feigel'man~\cite{FoFe01}. By means of the coupled Usadel equations for the hybrid system, they investigated the density of states as well as the parallel and perpendicular critical fields of the bi-layer as a function of the interface transparency. The asymptotics at high and low transparencies are accessible to analytical solutions while the results at intermediate transparencies were found numerically.

Here, we consider a different aspect of the properties of the hybrid system, namely the interplay between gapless superconductivity and the proximity
effect. In addition to the effect of the field on the
individual system, in the NS bi-layer it also affects the coupling. 
As we have seen
in the previous sections, an in-plane magnetic field gradually
suppresses the gap in the single-particle DoS. On the other hand, in
an NS structure, the proximity effect opens a gap in the DoS of
the normal layer. Thus, we expect the magnetic field to weaken the
proximity effect.

\begin{figure}[h]
\begin{center}
\epsfxsize=3.2in\epsfbox{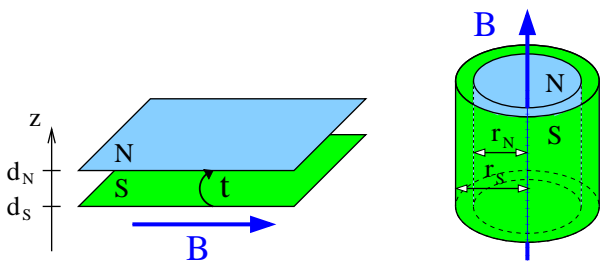}
\caption{\label{fig-ns}Sketch of the hybrid NS systems considered in the text in the planar and
    cylindrical geometry.}
\end{center}
\end{figure}

For simplicity, we consider here a hybrid system each consisting of a single N-
and S-channel (see Fig.~\ref{fig-ns}). I.e., neglecting the finite width,
the magnetic field does not influence the individual systems, and we
can study the effect on the coupling alone. The coupling between the layers is described by a tunneling Hamiltonian $\cH_T=\int d^2r \,(t\, \bar\Psi_N\Psi_S+{\rm h.c.})$, where $t$ is the tunneling matrix element (assumed to be spatially constant).
Thus, the effective action for the NS system consists of a sum of the actions 
of the individual systems, $S_N$ and $S_S$, and a coupling term that
in the weak tunneling limit can be linearized. I.e., the full action
reads (see e.g.~\cite{FLS})
\begin{eqnarray}
S\eq-\frac{\pi\nu_\Box}8\int d^2r\, \Tr\Big[D_N(\tilde\partial Q_N)^2-4i\epsilon\sph_3\otimes\scc_3Q_N+\nn\\
&&\qquad+D_S(\tilde\partial
Q_S)^2-4(i\epsilon\sph_3\otimes\scc_3-\hat\Delta\sph_2)Q_S-\nn\\
&&\qquad-4\gamma Q_NQ_S\Big],\label{gamma3}
\end{eqnarray}
where $\gamma=|t|^2\tau$ represents the transparency of the interface. Furthermore
$\tilde\partial=\partial-i{\bf A}[\sph_3,\,.\,]$, where the
appropriate gauge for the vector potential $\bA(\br)$ will be specified
shortly. In general, the
order parameter may be complex, $\hat\Delta=|\Delta|\exp[i\chi\sph_3]$.
Note that $\Delta$ here is the self-consistently determined order parameter. The presence of the normal layer leads to a renormalization of the order parameter. At weak coupling, however, the proximity induced suppression of the order parameter is small. Therefore, we concentrate on the much more pronounced effect on the quasi-particle DoS.

Subjecting the action~(\ref{gamma3}) to a saddle point analysis, one obtains the following coupled Usadel-like equations,
\begin{mathletters}
\begin{eqnarray}
D_N\tilde\partial(Q_N\tilde\partial Q_N)-[i\epsilon\sph_3\otimes\scc_3,Q_N]\eq\gamma[Q_S,Q_N],\nn\\
D_S\tilde\partial(Q_S\tilde\partial Q_S)-[i\epsilon\sph_3\otimes\scc_3\!-\!\hat\Delta\sph_2,Q_S]\eq\gamma[Q_N,Q_S].\nn
\end{eqnarray}
\end{mathletters}
As a simple guiding example, let us first consider the properties of the system in the
absence of a magnetic field, where the order parameter can be chosen
to be real. 
Now, in situations where the superconducting terminal is represented by a bulk system, the latter simply acts as a boundary condition for the normal region. However, in the present case, the single superconducting channel is itself heavily influenced by the contact with the normal region. As a result, Fominov and Feigel'man~\cite{FoFe01} have shown that  
a gap develops in the
normal region while in the superconductor quasi-particle states at
energies down to the size of the proximity effect induced gap are generated.

To see this explicitly, let us employ the Ansatz $Q_X=\cosh\hat\theta_X\sph_3\!\otimes\!\scc_3+i\sinh\hat\theta_X\sph_2$
with $\theta_X$ homogeneous ($X=N,S$). In this case,  the saddle point equations
reduce to
\begin{mathletters}
\begin{eqnarray}
-i\epsilon\sinh\hat\theta_N\eq\gamma\sinh(\hat\theta_N-\hat\theta_S),\nn\\
-i(\epsilon\sinh\hat\theta_S-\Delta\cosh\hat\theta_S)\eq\gamma\sinh(\hat\theta_S-\hat\theta_N).\nn
\end{eqnarray}
\end{mathletters}
If the two systems are decoupled, $\gamma=0$, the solution for the superconductor
at energies well below the gap, $\epsilon\ll\Delta$, reads
$\theta_S\approx i\pi/2$.
At weak coupling, setting $\theta_S=i\pi/2+\vartheta_S$ in the low-energy
regime and expanding the
equations above up to linear order in $\vartheta_S$ yields
$\coth\theta_N=\epsilon/\gamma$ and
$\vartheta_S={(\epsilon-i\gamma\cosh\theta_N)}/{\Delta}$.
Thus, at small energies, the density of states in the two layers is
given as
\begin{eqnarray}
\nu_N(\epsilon)\eq\nu_\Box\,\Re[\cosh\theta_N]=\left\{\begin{array}{ll}0&\epsilon<\gamma,\\
\nu_\Box\frac{\epsilon}{\sqrt{\epsilon^2-\gamma^2}}\quad&\epsilon>\gamma;
\end{array}\right.\\
\nu_S(\epsilon)\eq-\nu_\Box\,\Im[\vartheta_S]=\frac\gamma\Delta\nu_N(\epsilon).
\end{eqnarray}
As expected,
the superconductor induces an energy gap in the normal region of magnitude
$E_{\rm gap}^{(N)}=\gamma$. Furthermore, the contact with the normal
region leads to the appearance of quasi-particle states in the
superconductor at energies down to the proximity effect induced gap $E_{\rm gap}^{(N)}\ll\Delta$.
(Note that close to the singularity at $\epsilon=E_{\rm gap}^{(N)}$, the
approximations above are no longer valid.)

How do these characteristic features of the proximity effect change in 
the presence of a magnetic field? In the following, we will consider two
different geometries as depicted in Fig.~\ref{fig-ns}. In
Sec.~\ref{sub-plan}, a planar NS bi-layer is investigated.
Subsequently, in Sec.~\ref{sub-cylin}, we study a setup where the 
bi-layers are wrapped around a cylinder with the magnetic field
directed along the cylinder axis. While in the first case the
magnetic field effect is due only to the flux enclosed between the N- and S-layer,
here the system as a whole encloses magnetic flux which leads to
markedly different behavior. 

\subsection{Planar geometry}\label{sub-plan}

In the planar geometry, as before, the appropriate gauge is the London gauge
$\bA=-(Hz+c_0){\bf e}_y$. Here the constant $c_0$ is determined
through the condition that the supercurrent through a cross section of 
the bi-layer vanishes~\cite{FoFe01}:
\begin{eqnarray}
\int \! dz \;{\bf j}(z)={\bf j}_N+{\bf j}_S=0,
\end{eqnarray}
where ${\bf j}_X=n_X{\bf A}_X/m$. To a first approximation, $n_N=0$.
Thus, no supercurrent flows in the $N$ region and, therefore, the 
supercurrent in the $S$ region has to vanish as well, i.e.~$\bA_S=0$
which implies $c_0=-Hd_S$.

Using the same Ansatz as for the field-free case, one obtains
\begin{eqnarray}
-\frac{D_N}4(Hd)^2\sinh(2\hat\theta_N)-i\epsilon\sinh\hat\theta_N\eq\gamma\sinh(\hat\theta_N\!-\!\hat\theta_S),\nn\\
-i(\epsilon\sinh\hat\theta_S-\Delta\cosh\hat\theta_S)\eq\gamma\sinh(\hat\theta_S\!-\!\hat\theta_N),\nn
\end{eqnarray}
where $d=d_N-d_S$ is the distance between the two layers.

As pointed out earlier, being a single-channel
system, the superconductor alone does not feel the magnetic field. Again we are interested in the DoS at energies well below the gap,
$\epsilon\ll\Delta$. As in the field free case, an expansion in $\vartheta_S=\theta_S-i\pi/2$
leads to
\begin{eqnarray}
\kappa_N\sinh(2\hat\theta_N)+2i(\epsilon\sinh\hat\theta_N-\gamma\cosh\hat\theta_N)=0,\label{kappa-n}
\end{eqnarray}
where $\kappa_N=D_N(Hd)^2/2$. 
Furthermore, $\vartheta_S={(\epsilon-i\gamma\cosh\theta_N)}/{\Delta}$
as before.
Thus, in the two channel case, the magnetic field leads to a
suppression of the proximity effect. Eq.~(\ref{kappa-n}) shows that the closing of the induced gap is 
described by the AG theory, where the relevant parameter is 
$\zeta_N=\kappa_N/\gamma$. Therefore, one finds that the characteristic field
$\zeta_N(H_c)\equiv1$ causing the proximity effect induced gap to vanish is much weaker 
than the field necessary to drive the superconductor into the gapless phase (i.e.~taking into account the finite width of the individual layer!).

\subsection{Cylindrical geometry}\label{sub-cylin}

While in the planar geometry a non-vanishing supercurrent is
forbidden, in the cylindrical geometry a supercurrent can flow around the cylinder:
in contrast to the previous case, the system now encloses
magnetic flux.

Starting with a single superconducting layer, for the cylindrical geometry, the most convenient gauge to choose is
the symmetric gauge $\bA=\frac12\bH\times\br=-\frac12Hr{\bf
  e}_\varphi$, where we use cylindrical coordinates $(r,\varphi,z)$. Now the phase degree of freedom has to be taken into
account, and a more
general Ansatz for the matrices $Q$ solving the saddle point equations
is needed:
\begin{eqnarray}
Q_S=\cosh\hat\theta_S\sph_3\!\otimes\!\scc_3+i\sinh\hat\theta_Se^{in\varphi\sph_3}\sph_2,\nn
\end{eqnarray}
where $n\in\mathbb{Z}$, and the phase matches the phase of the order
parameter, $\hat\Delta=|\Delta|\exp[in\varphi\sph_3]$. Substituting this Ansatz into the saddle point equation, one obtains 
\begin{eqnarray}
\frac{D_S(n\!+\!Hr_S^2)^2}{2r_S^2}\sinh(2\hat\theta_S)+2i(\epsilon\sinh\hat\theta_S\!-\!\Delta\cosh\hat\theta_S)=0.\nn
\end{eqnarray}
Thus, even in the absence of the NS coupling, the magnetic field
affects the properties of the superconductor. 
The integer $n$ has to be chosen such
that it minimizes $|n+Hr_S^2|$, and, therefore, by increasing the magnetic
field, the parameter governing the gap suppression,
\begin{eqnarray}
\zeta_S(H;n)=\frac{D_S(n+Hr_S^2)^2}{2r_S^2\Delta},\nn
\end{eqnarray}
varies periodically between $0$ and
$\zeta_S^{\rm max}=D_S/(8r_S^2\Delta)=(\xi/d_S)^2$, where $d_S$ is the diameter
of the cylinder. The
size of the energy gap is
$E_{\rm gap}=\Delta(1-\zeta_S^{2/3})^{3/2}$, and the superconductor enters the gapless phase when 
$\zeta_S=1$. This condition can only be fulfilled if $d_S<\xi$. 

As pointed out earlier, the magnetic field not only suppresses the energy gap, but also renormalizes the order parameter $\Delta$. I.e.~the order parameter in the formulae above has to be determined self-consistently. At $T=0$, the self-consistency equation can be cast in the form~\cite{maki}
\breakon
\begin{eqnarray}
\ln\left(\frac\Delta{\Delta_0}\right)\eq\left\{\begin{array}{ll}{-\frac\pi4\zeta_S} & {\zeta_S\leq1,}\\{-{\rm arcosh\,}\zeta_S-\frac12\left(\zeta_S{\,\rm arcsin\,} \zeta_S^{-1}-\sqrt{1-\zeta_S^{-2}}\right)\qquad}&{\zeta_S>1.}\end{array}\right.
\end{eqnarray}
\breakoff

\begin{figure}[h]
\begin{center}
        \epsfxsize=3in\epsfbox{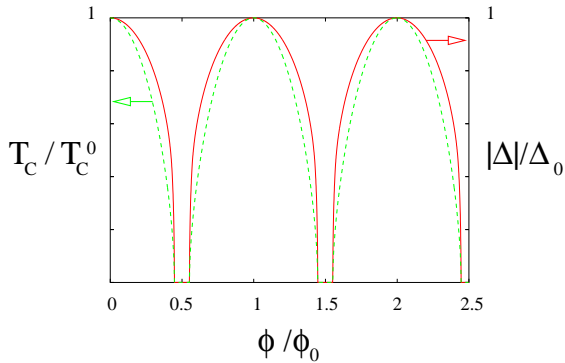}
\caption{\label{fig-exp} The order parameter $\Delta$ and the
  transition temperature $T_c$ as a function of the magnetic flux
  $\Phi=\pi Hr_S^2$ threading the cylinder.}
\end{center}
\end{figure}

\begin{figure}[h]
\begin{center}
\epsfxsize=3in\epsfbox{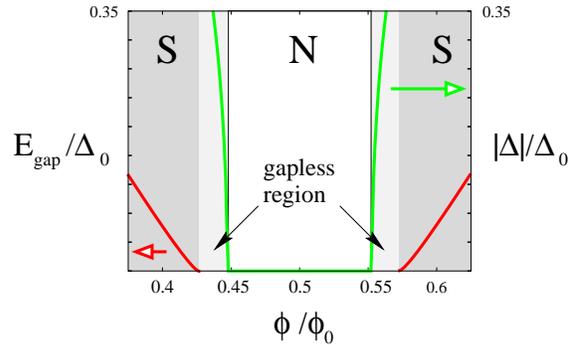}
\caption{\label{fig-gapless} Energie gap $E_{\rm gap}$ and order
  paramter $\Delta$ in the vicinity of $\Phi/\phi_0=1/2$.}
\end{center}
\end{figure}

\noindent
Thus, the periodic modulation of $\zeta$ also leads to a periodic modulation of the order parameter.
Similarly, the transition temperature $T_c$, which obeys~\cite{maki}
\begin{eqnarray}
\ln\left(\frac{T_c}{T_c^0}\right)=\psi\left(\frac12\right)-\psi\left(\frac12+\frac\kappa{2\pi T_c}\right),
\end{eqnarray}
is a periodic function of the applied field as first observed by
Little and Parks~\cite{little-parks}. Furthermore, in small rings
superconductivity is completely suppressed in a certain range of
magnetic fields around half-integer flux quanta threading the
cylinder~\cite{deGennes}. Only very recently it has been possible to
manufacture small enough cylinders, where this prediction could be
verified experimentally~\cite{newexp}. 
Fig.~\ref{fig-exp} shows the
order parameter at $T=0$ and the transition temperature for a cylinder
with $d_S<\xi_0$ (where $\xi_0$ is the coherence length at $H=0$),
plotted against magnetic flux $\Phi/\phi_0$, where $\Phi=\pi H r_S^2$. As expected, both
vanish around half-integer flux-quanta while they reach their
unperturbed maximal values at integer flux quanta. The region around
$\Phi/\phi_0=1/2$ is magnified in Fig.~\ref{fig-gapless}, where the
energie gap and the order parameter are plotted. The system shows a
cross-over S $\to$ gapless S $\to$ N $\to$ gapless S $\to$ S
by increasing the magnetic field.


Now adding the normal layer, we again concentrate on energies much
smaller than the gap. With  the same approximation used earlier, the
equation for the normal region reads
\begin{eqnarray}
\frac{D_N(n\!+\!Hr_N^2)^2}{2r_N^2}\sinh(2\hat\theta_N)+2i(\epsilon\sinh\hat\theta_N\!-\!\gamma\cosh\hat\theta_N)=0.\nn
\end{eqnarray}
Once again, the mean-field equation assumes the form of an AG equation with the parameter
$\zeta_N(H;n)=D_N(n+Hr_N^2)^2/(2r_N^2\gamma)$. Comparing the two
values $\zeta_S$ and $\zeta_N$, we find that $\zeta_N^{\rm
  max}\gg\zeta_S^{\rm max}$, i.e.~the proximity gap is suppressed
before the superconductor itself would enter the gapless regime. At
the same time the solution for
the superconductor takes the form
\begin{eqnarray}
\vartheta_S=\frac{\epsilon-i\gamma\cosh\theta_N}{\Delta(1-\zeta_S)},\nn
\end{eqnarray}
yielding $\nu_S=\nu_\Box\gamma/\Delta(1-\zeta_S)^{-1}\cosh\theta_N$.
I.e.~the combined influence of the presence of the normal region and
the magnetic field leads to an enhanced density of states at low
energies in the superconductor. 

This concludes our discussion of the mean-field properties of thin disordered NS hybrid systems. Taking into account the influence of fluctuations, it is straightforward to see that the low-energy properties of the quasi-particle states are dictated by the same theory obtained in the previous section.
Here, we assume the disorder in the N- and S-channel to be
uncorrelated which violates $\cP_z$-invariance. Thus, the gapless
hybrid system is described by symmetry class C.


\section{Conclusions}\label{sec-conclusions}

To conclude, we have cast the properties of a disordered thin superconducting film subject to a parallel
magnetic field in the framework of a statistical field theory. In the 
mean-field approximation, known results from the AG theory~\cite{AG} are 
recovered. The same
phenomenology applies to diffusive films as well as films with
columnar defects. In the diffusive case, we have shown that -- within
the gapped phase -- taking into account inhomogeneous instanton
solutions of the saddle point equation, the hard gap is
destroyed. By analogy with the magnetic impurity problem~\cite{LaSi00}, 
exponentially 
small tails within the gap region appear. The same is to be expected
for the columnar defects. (For
$M>2$ the coupling between the different subbands complicates the
analysis. However, the general behavior should not be affected qualitatively.)

Within the gapless phase, the Berry-Robnik symmetry phenomenon leads
to different low-energy properties. As confirmed by numerics, for the
diffusive film, the DoS vanishes quadratically for $\epsilon\to0$
(class C) as one might expect for superconducting systems where $\cT$-invariance is lifted. However, in the presence of only columnar defects, the DoS at
small energies is linear in $\epsilon$ (class CI), a behavior characteristic of systems which possess time-reversal invariance. Although
the $\cP_z$-symmetry cannot prevent the gradual destruction of
superconductivity by the magnetic field, some compensation for the
$\cT$-breaking is still effective. 

In NS bi-layers, we have shown that the coupling between the two systems leads to (i) an
energy gap $E_{\rm gap}^{(N)}$ in the DoS of the normal layer, and (ii)
a finite density of states in the superconductor at energies
$E_{\rm gap}^{(N)}<\epsilon<\Delta$. In this geometry, a parallel magnetic field suppresses 
the induced proximity gap $E_{\rm gap}^{(N)}$. The characteristic field $H_c(N)$
determining the occurrence of the gapless phase is greatly reduced as
compared to the field $H_c(S)$ that drives the superconductor into the gapless 
phase, being roughly $H_c(N)/H_c(S)\approx({E_{\rm gap}^{(N)}/\Delta})^{1/2}$.
In a cylindrical geometry, the energy gap shows a periodic modulation
with the magnetic field reminiscent of the Little-Parks effect: if the cylinder encloses multiples of the
flux quantum $\phi_0$, this can be compensated by the phase of the order
parameter. Thus, the variation of the energy gap is determined by
the effective field $H_{\rm eff}={\rm
  min}_{\,n\in\mathbb{Z}}|H+n/r_S^2|$. The gapless phase can only be
reached in sufficiently small systems, where the diameter of the cylinder fulfills the relation $d_S<\xi$.


{\sc Acknowledgments:} We would like to thank A.~Altland,
B.L.~Altshuler, D.E.~Khmel'nitskii and A.~Lamacraft for valuable discussions.


%
\end{multicols}

\begin{references}
\bibitem{fn1}
As in Ref.~\cite{MAA01}, it can be
shown that an asymmetry of the confining potential or the presence
of $z$-dependent scattering destroy any unusual phenomena associated with 
the Berry-Robnik symmetry effect~\cite{BeRo86}.
\bibitem{fn0}
In this limit, the supermatrix fields $Q$ can assumed to be diagonal in the subband space which simplifies the subsequent calculations. However, even in the opposite limit, the dominant contributions stem from diagonal $Q$, and the results are not expected to change qualitatively.
\bibitem{fn2}
One might expect that taking into account the phase of the order
parameter would change the situation, but we checked that this is not
the case.
\end{references}
\end{document}